\documentclass[usenatbib]{mn2e}
\usepackage{graphicx, amsmath, amssymb}

\begin{document}
\topmargin-1cm

\title{The geometry of the filamentary environment of galaxy clusters}

\author[Noh and Cohn]{Yookyung Noh${}^{1}$ and J.D. Cohn${}^2$\\
${}^1$ Department of Astronomy and Theoretical Astrophysics Center, University of California, Berkeley, CA 94720\\
 ${}^2$ Space Sciences Laboratory and Theoretical Astrophysics Center, University of California, Berkeley, CA 94720}

\date{\today}

\pubyear{2010}
\maketitle

\begin{abstract}
We construct a filament catalogue using an extension of the halo based filament finder of \citet{Zha09}, in a $250$ Mpc/h side N-body simulation, and study the properties
of filaments ending upon or surrounding galaxy clusters (within 10 Mpc/h).
In this region, the majority of filamentary mass, halo mass, and galaxy richness 
centered upon the cluster tends to lie in sheets, which are not always coincident.
Fixing a sheet width of 3 Mpc/h for definiteness, we find the sheet orientations and
(connected) filamentary mass, halo mass and richness fractions relative to the surrounding sphere.
Filaments usually have one or more endpoints outside the sheet determined by
filament or halo mass or richness, with at least one having a large probability to
be aligned with the perpendicular of
the plane.
Scatter in mock cluster mass measurements, for several observables, 
is often correlated with the observational direction relative to these local sheets, 
most often for
richness and weak lensing, somewhat less for Compton decrement, and least often for 
velocity dispersions.  The long axis of the cluster also tends to lie in the sheet and
its orientiation relative to line of sight also correlates with mass scatter.

\end{abstract}
\section{Introduction}
Large scale structure in the universe forms a cosmic web \citep{ZelEinSha82, ShaZel83, Ein84, BonKofPog96}, evident in the universe's dark matter, halo, galaxy, and gas distributions. 
The richness of the cosmic web is evident when one has
sufficient statistics and  resolution (numerically) or sensitivity (observationally) to see beyond the densest structures, correspondingly there has been a wealth of
study of its properties.  Examples include characterization of average properties (e.g. see 
\citet{Sch98} for one early review, \citet{Sha04, Wey10,Sha10} for some more recent papers
and references within), identifying
the web in observations and simulations (e.g., \citet{Bha00, PimDriHaw04, Fei08, Sou08a, BonStrCen09, WayGazSca10, BonStrCen10, Cho10, MeaKinMcC10, MurEkeFre10, SouPicKaw10}, tracing its
relation to initial conditions (e.g., \citet{ShaHabHei09}) and comparing filamentary environments and properties of galaxies within them (spin, shapes, alignments and more: \citet{Lee04, Lee08, AltColCro06, Ara07b, Dol06, PanSom06, Fal07, RagPli07,Hah07a, Hah07b, 
PazStaPad08, BetTru09, Gay09, Sch09, Zha09, HahTeyCar10, JonvanAra10,Wan10}). 
Cluster alignments and formation, presumably or explicitly along filaments, have also been studied, e.g. \citet{vanBer96, Spl97, Col99, ChaMelMil00, OnuTho00, Fal02, van02, HopBahBod04, BaiSte05, Fal05, KasEvr05, LeeEvr07, Lee08, PerBryGil08, CosSodDur10}, and
several observed systems with filaments have been analyzed in detail, 
some examples are found in \citet{PorRay05, Gal08, Kar08, Tan09}.  Numerous methods for identifying filaments, suitable for different applications, have been proposed
(for example \citet{BarBhaSon85, MecBucWag94, SahSatSha98, Sch99, ColPogSou00, She03, Pim05a, Pim05b, Sto05, NovColDor06, Ara07a, Col07, vanSch07, Sou08b, StoMarSaa08, For09, GonPad09, Pog09, SouColPic09, StoMarSaa09, Wu09, Gen10, MurEkeFre10, Sha10,Sou10,WayGazSca10}), see \citet{Zha09, AravanJon10} for
some comparisons of these.   Analytic studies of filaments
include estimates of their multiplicity
\citep{Lee06, She06}, anisotropy (e.g.\citet{LeeSpr09}), the
merger rates of halos into them \citep{SonLee10}
and properties in non-Gaussian theories \citep{SimMagRio10}.  

Galaxy clusters (dark matter halos with mass $M \geq 10^{14} h^{-1} M_\odot$) are of great interest for many 
reasons, in part because of their sensitivity to cosmological parameters, but also as hosts of the most massive galaxies in the
universe, as environments for galaxy evolution more generally, and as the largest
virialized objects in the universe with correspondingly special astrophysical
processes and histories (for a review see e.g. \citet{Voi05}).  
Galaxy clusters tend to lie at nodes of the
cosmic web, with matter streaming into them from filaments (e.g. \citet{vanvan93, DiaGel97, Col99}). 
Although the universe is isotropic and homogeneous on large scales, around any individual cluster there will be directionally dependent density fluctuations due to the condensation of filamentary and sheetlike matter around it.
Our interest here is in characterizing this nearby (within 10 Mpc/h) filamentary environment of galaxy clusters.  This environment feeds galaxy clusters and is also unavoidably included for many observations of the
cluster at its center.  This correlated environment
is one source of the observationally well known ``projection effects,'' which have plagued optical cluster finding
starting with \citet{Abe58} and later
(e.g. \citet{Dal92,Lum92, vHaarlem97,Whi99}),  cluster weak lensing, e.g., \citep{RebBar99,MetWhiLok01,Hoe01,PutWhi05,Men10,BecKra10},
cluster 
Sunyaev-Zel'dovich \citep{SunZel72, SunZel80} (SZ) flux measurements, e.g., \citep{WhiHerSpr02,HolMcCBab07, Hal07,ShaHolBod08} and cluster velocity dispersions, e.g., \citep{Cen97,Tor97,KasEvr05,Biv06}.
The environments of clusters have been studied within several contexts and
using several methods, e.g. galaxy and
dark matter density around clusters \citep{Wan09, Pog10}, filamentary 
growth (e.g. \citet{van06}) around clusters, filamentary counts \citep{ColKruCon05, AravanJon10, AraShaSza10}, in particular the
geometry and properties of superclusters, e.g., \citet{ShaSheSah04,Bas06, Wra06, CosSodDur10},  and
the cluster alignment studies such as mentioned above.

Here we describe our findings on local cluster environments obtained by implementing the
halo-based filament finder of \citet{Zha09} in a
high resolution N-body simulation.  After refining the finder slightly for our purposes, we obtain a filament catalogue, and consider those filaments connected to or in the vicinity of galaxy clusters.  
Our work is most closely related to that of \citet{ColKruCon05, AravanJon10}.
They used simulations to measure  counts of filaments (found via different algorithms) ending upon clusters and average filamentary profiles and curvature. 
We go beyond these to measure the statistics of the local geometry of filaments around their cluster endpoints.  Related studies of filament geometry, particularly for
superclusters are found in e.g.,
\citet{AravanJon10, AraShaSza10}, the former also discuss the tendency of filaments
around voids and clusters to lie in sheets. 
We find that most of the filamentary (and halo) material in a 10 Mpc/h sphere around clusters lies in a plane, presumably the one from which the filaments collapsed, 
and investigate different ways of defining such a plane's orientation.
Many measures of cluster masses include the cluster environment and as a result
scatter the mass from its true value.  In mock observations on simulations, we find that
line of sight dependent scatter in measured cluster masses, for
several methods,
is often correlated with the angle between the line of
sight and these locally defined planes.

 In \S 2 we describe the simulations, mock observations, and filament finder.
In \S 3 we describe the statistical properties of the filaments and matter distribution
around clusters, in \S4 we consider the geometry of the filament, mass and richness 
distributions within 10 Mpc/h of each cluster, focussing particularly on planes
 maximizing
these quantities, in \S 5 we compare scatter in cluster masses to orientation of observations with these
planes, and in \S 6 we conclude.

\section{Simulations and Methods}
\subsection{Simulation}
We use a dark matter only simulation, 
in a periodic box of side $250 $ Mpc/h with $2048^3$ particles evolved using the TREEPM \citep{TreePM} code, and provided to us by Martin White.  
It is the same simulation as used in \citet{WCS} (hereafter WCS),  which can be consulted for details beyond those found below.
The background cosmological parameters are $h=0.7$, $n=0.95$, $\Omega_m=0.274$, and $\sigma_8=0.8$, in accord with a large number of
cosmological observations.   The simulation has outputs at 45 times equally spaced in $\ln(a)$ from $z=10$ to $0$.   We focussed on $z=0.1$, in part to allow
comparison with observational quantities in \S 5.
Halos are found using a
Friends of Friends (FoF) halo finder \citep{DEFW}, with linking length $b=0.168$ times
the mean interparticle spacing.  Masses quoted below are FoF masses.

Resolved subhalos in this high resolution simulation are
of importance for the observational comparisons in \S 5, and for measurements of
galaxy properties around and in the clusters.  Subhalos are
found via FoF6d \citep{DieKuhMad06}, with the
specific implementation as described in the appendix of WCS.  The subhalos
correspond to galaxies with luminosities  $\geq 0.2 L^*$ at $z=0.1$\footnote{Approximately -18.5 in $r$ band, see WCS for more discussion.},
and match observations as described in WCS.  The halo and subhalo catalogues and
dark matter particles can be combined to produce mock observations for six 
cluster mass measures.  These are (see WCS for specifics and tests of the 
catalogue):  two
richnesses (one using the MaxBCG \citep{maxBCG} algorithm based
upon colors\footnote{Color assignments are estimated
using the prescription of \citet{SkiShe09} with evolution of \citet{SP1,SP2,SP3}.}, and one based
upon spectroscopy \citep{YanMovan08}),  SZ flux or Compton decrement
(flux within an annulus of radius $r_{180b}$, the radius within which the
average mass is greater than or equal to 180 times background density), 
weak lensing (using an SIS or NFW model to assume a cluster lens profile and then fitting for a velocity dispersion and then mass), and two velocity dispersions 
(one based on a simple
$3-\sigma$ clipping, the other on a more complex method using phase space information
to reject outliers and calculating mass using a measured harmonic radius
as well, based on methods of \citet{denHartog, Biv06, Woj07}); more detail is in WCS.
We will use the mass measurements by WCS via these methods, taking cylinders
of radius $r_{180b}$ when a radius choice is required.  Just as in that work, 
lines of sights for clusters are removed a more massive cluster has its center
within this radius along the observational line of sight.

\subsection{Filament finder}
We find filaments using an extension of the method described in \citet{Zha09}.
They identify filaments as
bridges in dark matter halos above a threshold halo mass overdensity, of  length up to 10 Mpc/h.
It  is analogous to the spherical overdensity finder for clusters, where the cluster radius is taken to be that where the average density around the central point drops below some threshold; here the filament radius is where the average
density along the cylinder axis drops below some threshold.
Just as there are many different halo finders, 
there is no unique filament finder or definition.  This finder is but one of many
different ones present in the literature, which are not only based upon such
bridge-like definitions, but also include finders constructed around
filtering procedures, 
potential or density gradients, dynamical information, and more
(see \citet{Zha09} for some comparisons between their finder and others).
Even for a given filament finder, catalogues often
must be specified by the finder parameters as well (e.g. smoothing length for
density or potential based finders, unbinding criteria for dynamically based
finders, etc.).  We use the parameters as given in \citet{Zha09}.

The algorithm of \citet{Zha09} is as follows: halos are ordered most to least massive. 
All halos with mass $\geq 3 \times 10^{10} h^{-1} M_\odot$ are included\footnote{This is the minimum mass used by \citet{Zha09} converted (see \citet{Whi01}) to our FoF definition.}, mass in the following only refers to that in halos with this mass or above. Starting with the most massive halo (``node''), all halos within 10 Mpc/h but at least 3 Mpc/h away in radius (or $r_{200c}$, if greater) are considered as potential endpoints.  For each potential endpoint, the cylinder radius is varied, up to 3 Mpc/h, to get the highest overdensity of halo matter in the cylinder between the node and potential endpoint.  This maximum density is then compared to a minimum overdensity (5 times background matter density in halos), and if over this minimum, this endpoint and its radius are kept.  If no potential endpoints have a halo mass density for their filament greater than the minimum overdensity
then the algorithm moves to the next node.  Once all such maximal
filaments are found for a given node, the filament with the largest density is kept.  The filament is then truncated:  its new endpoint is the most massive halo within it which
has at least 3 other halos between in and the central node, and 
which is at least 3 Mpc/h away from the central node.  
All filament members are then removed from the list of potential future filament
members or endpoints around any node.  The endpoints are not removed from the list of possible endpoints for other nodes, but are removed from the list of possible
endpoints associated with this node.  This procedure is repeated until no more
new filaments are found around the node.

As this procedure frequently
produces many more filaments than were evident by eye around clusters (sometimes over thirty around a single cluster), we incorporate a growing and merging procedure
as well.  After finding the filaments of maximum density around a given node, 
we grow out the filament radii until the average mass density in halos within the cylinder stretching to the filament endpoint drops to less than the minimum overdensity, or the maximum 3 Mpc/h radius is reached. Halos lying in two or more such extended filaments
are assigned to the one whose axis is closest.  
Filament endpoints with length $\ell$ and a perpendicular distance $d_\perp$ to another (longer) filament's axis such that $d_\perp /\ell < 3/10$ (the maximum width/maximum length in the algorithm) are merged into the longer
filament, unless the shorter filament's endpoint has
other filaments extending out of it.  (This allows filament radii $>$3 Mpc/h.)
These new filaments are then given a central axis determined by the center of
mass of the filament; filaments whose endpoints do not have additional filaments extending out
of them and whose endpoints are within 25 degrees of each other
are merged.  This is done in order of closest to most distant pairs; if $>2$ filaments are within this range, the two closest are merged, then centers of mass are recalculated to see if the remaining filaments are within the minimum distance, and so on.

The resulting filaments are 
regions connecting halos with halo mass overdensity
at least 5 times the background halo mass density, and which are less than 10 Mpc/h long.  
The full catalogue at $z=0.1$ has  $\sim 30, 000$ filaments and $\sim 44, 000$ endpoints, 
with 45\%  of the halo mass fraction in filaments and 36\% of the halos (in number
fraction) in filaments.  
 60\% of the $\sim 1.2 \times 10^6$ halos
above the minimum mass cut are not either endpoints or in filaments, with the
most massive of these having $M= 2.6 \times 10^{12} h^{-1} M_\odot$.
\footnote{
Analytic estimates of filamentary mass fractions mentioned above (which use other filament
definitions)
are not directly comparable because
the latter are based upon total mass; mass in halos above our minimum is
only 40\% of the mass in the box at $z=0.1$.}

Several of the other finders produce filaments which can extend well beyond our
10 Mpc/h cutoff (e.g. \citet{ColKruCon05} found filaments out to 50 Mpc/h, 
even longer ones have been found by e.g, \citet{GonPad09}), some have
restrictions on filament nodes (e.g. \citet{ColKruCon05} filaments end only on clusters).
Our catalogue has straight filament segments $\leq 10 $ Mpc/h in length, built out
of dark matter halos above some minimum overdensity which emanate from clusters and other endpoints.  Longer filaments could presumably be constructed as
chains of our shorter ones, augmented by a condition on how much a
filament can bend before it is considered instead to be two separate filaments meeting
at a node.    The length restriction of our finder also affects breakdowns into mass fraction in filaments, nodes, and so on, as some of our nodes will instead be filament members if the filaments are extended this way.

The work most similar to ours in focus, studying clusters as filament endpoints, is in \citet{ColKruCon05}, some related results are also found and compared in
\citet{AravanJon10} (see also \citet{SouPicKaw10}, who found a cluster as an intersection of filaments in observational data). \citet{ColKruCon05} found
filaments by looking for matter overdensities by eye between cluster endpoints and measured a wide range of
filament statistics, including the number of filaments per cluster as a function of mass, 
stacked filament profiles, length distributions, and the fractions of cluster pairs
connected by filaments.  \citet{AravanJon10} found filaments
using a Multiscale Morphology Filter (see their paper for details) and 
considered similar quantities to \citet{ColKruCon05}, and in addition introduced a classification for filaments.

\section{Statistics of filaments around clusters}

Our finder is well suited to characterize the local environment of clusters, our target of study here.  Of the 
242 clusters ($M\geq 10^{14} h^{-1} M_\odot$) in our box, 226 are also
nodes, with $\sim 1700$ filaments.  We restrict to these clusters below.
The other 7\% (16) of the
clusters are within
filaments linking two more massive clusters, in addition, 29 of the clusters have a cluster
within a filament, and 41 cluster pairs are within a 10 Mpc/h radius of each other.
We use the term ``connected'' filamentary mass to refer to halo mass within
a filament connected directly to a cluster, up to and including its other endpoint.
\footnote{The finder, even with modifications, still produced some configurations
which we modified with post-processing.  For example,
sometimes a filament would be found with a large ``gap'' in the center, 
where the gap is due to a previously found filament between two other clusters which
crosses the region.  Even with this gap, the new filament is above our
overdensity threshold.  As the previous and new filaments seem joined and perhaps one object,
we added all the mass (within 10 Mpc/h) of any previously found filament which came
within 3Mpc/h to the connected filamentary mass of the cluster,
this happened for $<10$ of our $\sim$7000 filaments.}
In addition to connected filaments around a cluster, within the 10 Mpc/h sphere
we will also
consider all filaments and their endpoints, all halos
above our minimum mass of $3\times 10^{10} h^{-1} M_\odot$, and all galaxies.  

In the 10 Mpc/h spheres surrounding clusters, 
connected filaments constitute $\sim 70\%$ of the halo mass on average, but
with a very broad distribution of values for individual clusters.  %plotfilmass,G.filfracs
A line passing through the 10 Mpc/h shell centered on a cluster will hit one of the original connected filament
cores (from the first step of our algorithm) about 10\% of the time on average, and one of the grown and merged
filaments closer to $\sim 30$\% of the time, with a wide spread as well.  %**fullongang**
%**proj_heal_ncov.c in proj_rot_solidang_v2, **fullang**
All (not only connected) filaments in this sphere contain closer to $\sim 90\%$ of the halo mass, 
with much less cluster to cluster scatter.  (These additional filaments connect
two other halos, which may or may not lie within the sphere themselves, rather
the cluster and another halo.)
In 10 Mpc/h spheres around 10,000 random points, in comparison, the filaments have a halo mass fraction ranging from 60\% to 95\%. %  **plotfilmass** average 67\%

\begin{figure}
\begin{center}
\resizebox{3.5in}{!}{\includegraphics{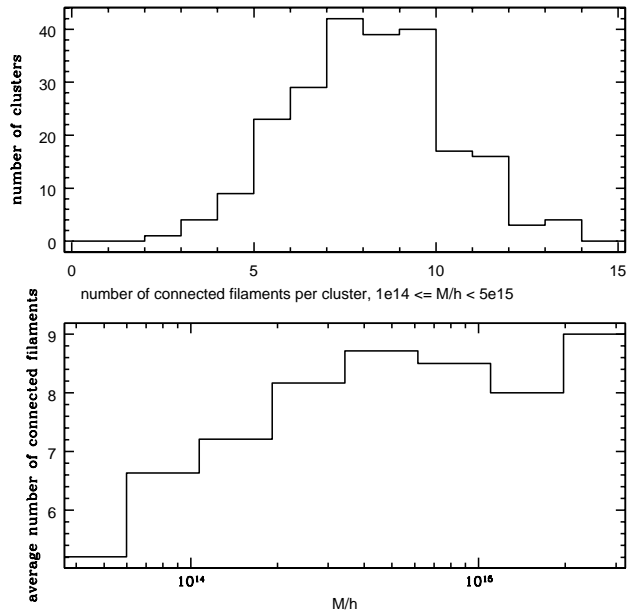}}
\end{center}
\caption{
 Top: distribution of number of filaments per cluster (halos with $M \geq 10^{14}h^{-1}M_\odot$). Bottom: number of filaments as a function of mass for all halos which are
filament endpoints.} %**plotsegs**
\label{fig:numfils}
\end{figure}

The distribution of number of connected filaments around clusters, with our finder, is shown at the top of  Fig.~\ref{fig:numfils}, clusters tend to have 7-9 filaments.  
We find more massive halos have more filaments ending upon them,
shown in Fig.~\ref{fig:numfils}, bottom,
just as found by \citet{ColKruCon05,AravanJon10} with their different finders.
In addition, connected filaments around clusters
tend to be shorter than their counterparts for the much less massive nodes.%**fillengths**

\begin{figure}
\begin{center}
\resizebox{3.5in}{!}{\includegraphics{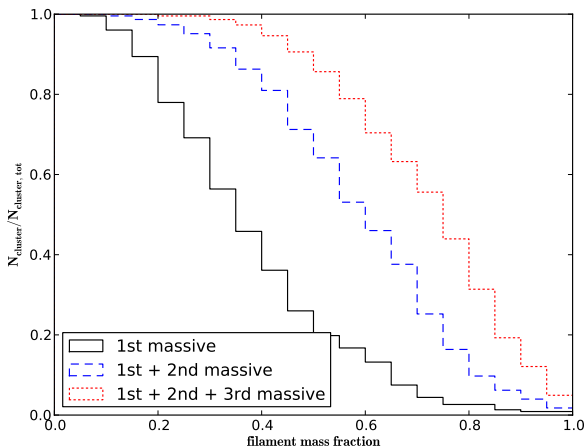}}
%\resizebox{3.5in}{!}{\includegraphics{fracsegmass.ps}}
\end{center}
\caption{Cumulative fraction of filamentary mass in 1 (solid line), 2 (dashed line) and 3 (dotted line) most massive cluster filaments, as fraction of the number of clusters.
For example, about half of the clusters have at least 
60\% of their connected filament mass
in their two largest filaments and at least $\sim 75$\% of their mass in their three largest filaments.  }
\label{fig:massivesegs}
\end{figure}

The large number of filaments found by the algorithm can be compared to 
a simplified picture where nodes are fed by a small number of filaments (e.g. 3 or fewer, 
\citet{Ker05, Ker09, Dek09}).
The mass fraction in the largest 2 or 3 filaments is substantial, leading to a partial reconciliation of these pictures, as seen in Fig.~\ref{fig:massivesegs}, that is 
about half of the clusters have at least $\sim 75$\% of their mass in their three largest
filaments.  

 %bigsumsegs in nodes.py
%**bigsegconts("segmassendfrac_14.0_15.5_rp1.5.dat") or without ends, higher by
%about 4-5% if use ends

More massive halos have more filaments around them, 
more matter in filaments, and more matter around them generally, and 
although the number of filaments for clusters can be quite large, a significant
fraction of the filamentary mass is found within the three largest filaments.

\section{Planar Geometry around clusters}
Filaments provide an anisotropic environment for galaxy clusters.  
Some approximate trends in the filamentary distribution are accessible
via the inertia tensor of its mass, even though
filaments are not expected to fill out an ellipsoid.
For our clusters,
the moment of inertia tensors tend to have two relatively large eigenvalues
and one smaller one (corresponding to axis ratios $a> b \sim c$, the classic
prolate cluster shape, 
there are many studies of cluster ellipticities, see e.g. \citet{JinSut02}.)
In comparison, for connected filaments attached to our clusters, the middle eigenvalue of the inertia tensor tends
to be smaller so that the filament distribution is ``flatter'' than the cluster it surrounds.\footnote{
The connected filament distribution becomes more and
more cylindrical with decreasing (well below $10^{14}h^{-1} M_\odot$) central halo mass, with the two largest eigenvalues tending to become equal, 
and the third becoming smaller and smaller.   One reason is that
lower mass halos are expected to
be within filaments, rather than to serve as endpoints; the algorithm used here
will tend to break these longer filaments up into more segments as mentioned earlier.}  %**evlowhigh** 
%**filiner**
The long axis of the cluster has a tendency to lie within the ``flat'' directions of
the filamentary distribution, and the eigenvector of the cluster's inertia tensor
that is perpendicular to the long and middle axes of the cluster (i.e. corresponding
to the largest eigenvalue) tends to align with
the corresponding direction of the filamentary inertia tensor.
(See also \citet{van06, Hah07b, Ara07b, PazStaPad08, AravanJon10}; as our nodes will sometimes be members of filaments in other finders, some of these
alignments are relevant filament member 
alignment discussed therein.)

Visual inspection of many of our clusters suggests that the majority of their filamentary
 mass lies within sheetline regions, presumably those from which they condensed (see for example some cases illustrated in \citet{AraShaSza10}, and discussion of different 
filament types in \citet{AravanJon10} and their ``grid'' and ''star'' configurations).

\begin{figure}
\begin{center}
\resizebox{3.5in}{!}{\includegraphics{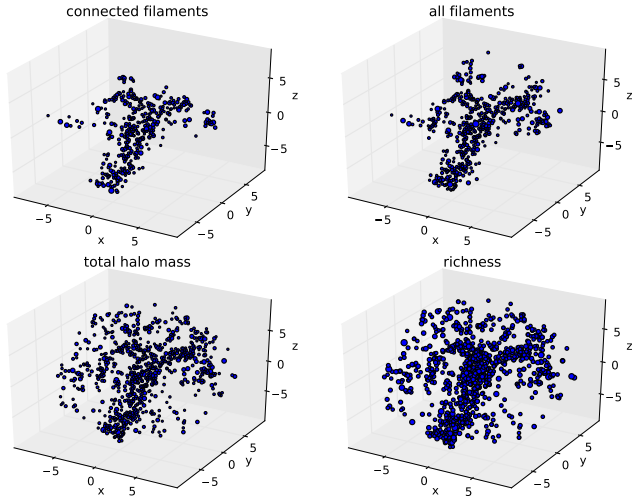}}
\end{center}
\caption{
Four types of objects used in constructing planes in a 10 Mpc/h radius sphere
centered on a $2.7 \times 10^{14} h^{-1} M_\odot$
cluster. Left to right, top to bottom are
halos in connected filaments, halos in all filaments, all halos above $3 \times 10^{10} h^{-1}M_\odot$ mass cut, and galaxies above $0.2 L^*$ cut (galaxies in cluster
are not shown).    
Point size is proportional to halo mass or, for richness, halo infall mass (which determines luminosity, see WCS).
About 84\% of the cluster's connected filament mass is in the connected filament 
plane.
}
\label{fig:fourplane}
\end{figure}

To quantify this planarity, we consider four definitions of planes,
regions extending $\pm 1.5$ Mpc/h above and below the 
central cluster and out to the edge of the local 10 Mpc/h sphere.
We choose their orientations (normals) so the planes contain the maximum of either 1)  connected filament mass, with 
extra constraints described below, 2) all filamentary mass including endpoints, 3) total halo mass, or 4) number of galaxies,
within the 10 Mpc/h sphere.  The connected filament mass plane 
has its normal chosen to be perpendicular to the axes constructed out of a pair 
filament endpoints; this definition has stronger correlations with observables (discussed 
later) than using pairs of connected filaments without their endpoints, or using the 
plane maximizing connected filament mass with no other constraints.
The mass in the plane (or richness, when using galaxies) does not include that
of the central cluster, as our interest is in the cluster's environment.
In Fig.~\ref{fig:fourplane} the objects used for these four choices of plane are shown
for a cluster of mass $2.7 \times 10^{14} M_\odot/h$.
It has about 84\% of its mass in the 
connected filament plane.
%G.fourplane(30)

These four planes tend to have similar orientations, with the
all filament and halo mass planes are most often aligned
(over 96\% clusters have these two normals within 30 degrees).
This is not surprising given the dominance of filamentary mass in the 10 Mpc/h sphere
around the cluster noted earlier.
For a given cluster,
the largest misalignment between any pairs of planes
tends to be %(3/4 of the time)
between its connected filament plane and one of the other
planes, which for 15\% of the clusters differs by another plane by
more than 60 degrees.%**cfilrest
For most clusters it thus seems that the
connected filaments are not as closely aligned with the other planes, which extend further out into the sphere.  %**cdots**
Plane pairs besides the closely aligned all filament and halo mass plane
have on average $5-10\%$ of the clusters mismatching by $>$60 degrees. %9\%!
%**cdots**
%1/5 of the clusters have %low66deg
%at least one pair of planes disagreeing by at least this much.%**cdots*

\begin{figure}
\begin{center}
\resizebox{3.5in}{!}{\includegraphics{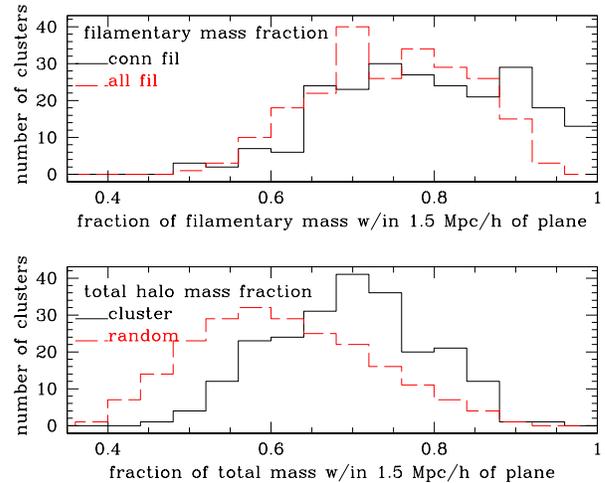}}
\end{center}
\caption{Top:
Fraction of connected filament mass in connected filament plane (solid) and 
fraction of all filamentary mass within all filament plane
(dashed), both in the fiducial 10 Mpc/h sphere around clusters.  
The normals to these planes are within 30 degrees for $\sim$80\% of %aff30 in **cdots*
the clusters.  
Bottom:  The fraction of total halo mass (above our $3\times 10^{10} M_\odot/h$ cutoff)
in the mass plane around clusters (solid), and its counterpart around 10,000 random points (dashed, rescaled to have the same number volume as cluster
histograms).  A large fraction of the filamentary mass and total halo mass in 10 Mpc/h spheres around clusters resides within this planar region containing $\sim$20\% of the volume.  
}
\label{fig:fracplane}
\end{figure}

The mass or richness fractions 
in these planes is significantly higher than the fraction ($\sim$1/5) of volume
which the plane occupies in the sphere.
The distribution of connected and total filament mass fractions, in
the corresponding planes, for our clusters is shown at top in
Fig.~\ref{fig:fracplane}, at bottom is the distribution for the total halo mass plane.
Also shown at bottom is the mass fraction for halo mass planes
constructed around 10,000 random points (rescaled to have the same area under
the curve), which is smaller on average than around the clusters.
The richness fraction, not shown, peaks slightly more sharply than the halo mass
fraction, but at a lower fraction ($\sim 60\%$).  For all plane definitions, 80\% of the
clusters have more than 60\% of their mass (or 55\% of their richness) 
in these planes;  about a quarter of the
clusters fell below this fraction for at least one plane definition.
Our choice of plane height, $\pm$ 1.5 Mpc/h to give 3 Mpc/h in total, 
was motivated by the characteristic scale
of cluster radii.  
We explored mass plane heights from 1-3 Mpc/h (total plane widths 2-6 Mpc/h),
and found that the total halo mass fraction scaled as $M_{\rm plane}/M_{\rm sphere} \sim {\rm height}^{1/4}$.  It would be interesting to understand this scaling
in terms of intrinsic filament profiles.  %**averad** 
%plotfinplanep**   has numbers, and 
%**plotfinplaneall** has everything.
%**cdots** lmin values gives these **minplanep()?? 

The clusters with large plane misalignments  (by $>$ 60 degrees) have
low mass or richness fractions, or larger mass within 3 Mpc/h of the normal of 
the connected filament plane (but outside of it)
almost twice as often as in the full sample (i.e. in $\sim$2/3 of the clusters with mismatched
planes).
%all3frac, all3fracall, not using pairs
The misaligned plane clusters have only slightly more often
a recent\footnote{Specifically, a satellite which has fallen into the cluster within the last time step, $\sim 600 Myrs$,  which had at the earlier time
at least 1/10 of the cluster's final mass at $z=0.1$.} merger or
a larger intrinsic cluster flatness (as measured by its inertia tensor), they %**flat--ellip<0.6%
were equally likely 
to have other clusters within 10 Mpc/h as in the full sample.
%**cdots**
 %alignlow.save 38% with mcyl>=10% in fil for align<0.5, vs. 13% for align>0.5
%cylsort.dat, segdistbig_14.0_etc, 

The connected filament plane's normal, similar to its counterpart for the
connected filament's inertia tensor,
tends to be aligned with its counterpart for
% **filiner**
the cluster's mass inertia tensor, and the cluster's long axis is likely to
lie in the filament plane.
The cluster galaxy positions, 
have an inertia tensor (setting mass to one) which appears uncorrelated with this plane, %upper left
%hand side,**gchist, gchistlong**
but restricting to more luminous ($>0.4 L_*$, see WCS for detail) galaxies gives
an inertia tensor whose ``most flat'' (perpendicular to eigenvector
for largest eigenvalue) direction prefers alignment with the normal to the connected
filamentary plane, and whose ``long'' axis tends to be within the filament plane.  
The galaxy velocity dispersions are larger along the connected filament
plane than perpendicular to it, 
but are more correlated (e.g. \citet{KasEvr05, WCS}) with the inertia
tensor of the cluster itself.
%**galcorrln, *galcorrlnlong**,**gchist in plots.sm%

\begin{figure}
\begin{center}
\resizebox{3.5in}{!}{\includegraphics{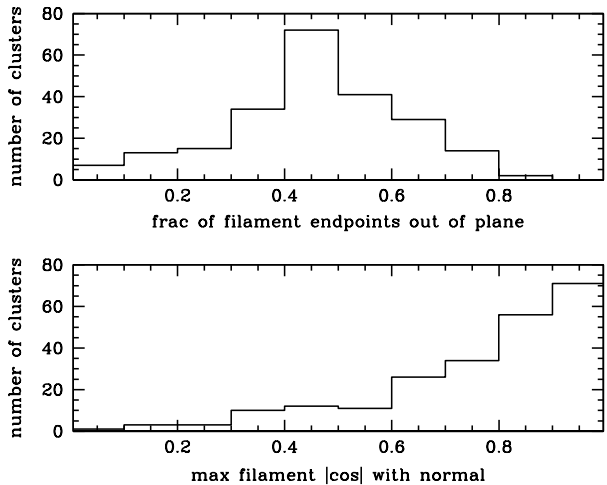}}
\end{center}
\caption{Top:  fraction of filament endpoints lying outside of connected filament plane-
many filaments do
not have their endpoints in this plane, even though a large fraction of mass is in
this plane (see Fig.~\ref{fig:fracplane}). Bottom: angle to normal of
connected mass plane, for filament closest to the normal;
at least one filament tends to be perpendicular to this plane. %plotfiloutboth**
%This is about the same probability for a random distribution of the median 
%number of cluster filaments to have a filament near any specific direction. 
The corresponding distributions for other planes are similar.
}
\label{fig:filout}
\end{figure}

Not all filaments lie in these planes.  Filamentary mass can extend outside
of the plane, as mentioned earlier, as can filament endpoints.  The fraction of
filament endpoints lying outside the connected filament plane
is shown in Fig.~\ref{fig:filout}, note that this does not preclude
a significant amount of the filament's mass lying within the plane.
%**plotfiloutboth**
There is also an increased likelihood for at least one endpoint to lie perpendicular to the
connected plane, as shown in Fig.~\ref{fig:filout} bottom.% but note**plotfiloutboth(0)
  The distributions
in Fig.~\ref{fig:filout} are similar for the other plane choices.
About 1/10 of the clusters have more than $\sim$3\% of %3.5 actually
their connected 
filamentary mass within a 3 Mpc/h radius of the normal to their plane but
above or below the plane itself, which we refer to as perpendicular filaments
below.  In addition 10 clusters have
over 15\% of their mass in a region within 6 Mpc/h radius of the normal, but
outside the connected plane.
%**perpmass in plots.sm**, makes cylsort
% in **cdots**

As noted earlier (Fig. ~\ref{fig:massivesegs}), 
the two most massive connected filaments often do possess a large fraction of the connected filament
mass.  The plane defined by these two filaments coincides with the connected
filamentary mass plane almost half % 40\% 
of the time.\footnote{We thank G. Jungman for asking
us to measure this.}%**plotbig_minplane**
For 1/3 of the clusters, however, less than half of the connected filament
planar mass comes from these two most massive segments.
%plotbig_minplane
% plotbig_minplane window 2 2 2 2 cumulative plot.
So although the two most massive segments have a preponderance of filamentary mass (Fig.~\ref{fig:massivesegs}), their large mass is
not wholly responsible for the dominance of planar structure.

The persistence of the locally defined planes to larger radii can be studied by 
fixing the plane height and orientation, and extending the plane out into a region
of 20 Mpc/h in radius, and calculating the fractional mass in this larger plane
within the larger sphere.  The plane volume fraction of the sphere volume drops by
about one half compared to its value in the 10 Mpc/h sphere, but the (all) filamentary mass, halo mass and  richness fractions in
their respective planes drop by even more, by a factor of $\sim 40\%$.
There are filamentary, mass or richness planes in this larger sphere of the
same $\pm 1.5$ Mpc/h width which have
more of the filamentary, mass or richness in them (and usually more than 1/2
the mass fraction of those defined within 10 Mpc/h).
These 20 Mpc/h filament and mass planes differ from their counterparts at
10 Mpc/h by over 30 (60) degrees one half (one quarter) of the time, with
slightly smaller fractions for the corresponding richness plane. 
 %mass20

We did not find a more useful measure of isotropy in the plane (i.e. in the angular
direction), 
although the moment of inertia tensor can indicate how much the planar geometry
tends to cylindrical (related questions have been explored
when classifying filaments, e.g.\citep{AravanJon10} note a ``star'' geometry for
sets of filaments).  One possible consequence of isotropy, or its lack, in the
plane will be discussed in the next section on mass measurements.

In summary, as has been known, the mass around clusters tends to lie in filaments, 
which themselves tend to lie within sheets.  
We have taken a set sheet width
centered on the cluster and maximized different quantities (filament mass, connected
filament mass, total halo mass and galaxy richness) within a
10 Mpc/h sphere around each cluster.  The resulting planes are not always aligned:
the all filament plane and all halo mass plane are most likely to be
aligned, and the largest disagreement between planes for any cluster is most likely to
be between the connected filament plane and another plane.  The long axis of the
cluster tends to lie in the plane as well.
Often
a perpendicular filament is also present relative to the plane, with others also partially extending out of
the sheet.
The rough cartoon of the filament shape around clusters is a planar structure
with a few filaments sticking out, with a tendency for at least one filament to be close to the plane's normal direction.

\section{Correlated mass scatter with local filamentary planes}
There are observable consequences of the filaments surrounding galaxy clusters:
most cluster observations, aside from X-ray \footnote{X-ray structure might have some correlation as well, inasmuch as X-ray substructure is
related to filaments which provide the cluster's infalling material.}, will tend to include
some of the cluster environment as well as the cluster itself. 
We saw above that the majority of the clusters have a preferred direction in their
local (10 Mpc/h radius) environments, with a large fraction of their surrounding
(connected or all filamentary, or total halo) 
mass or richness lying in a 3 Mpc/h sheet. The relation of this local 
structure to observables can be studied by using the mock observations
described in WCS.  In that work, 
cluster masses were measured along 96 lines of sight, 
using six methods mentioned earlier:
two richnesses, Compton decrement, weak lensing, and two velocity dispersions.
For individual clusters, WCS found correlated
outliers in the mass-observable relation along different lines of sight. 
(It should be noted that Compton decrement and weak lensing both can have
significant contamination from beyond the 250 Mpc/h path measured within the box,
so correlation with the local environment is likely smaller than found in
WCS and below.)
Some connection with environment or intrinsic properties is seen:
for the 8\% of cases where at least two observables had a large ($\geq 50\%$) deviation in mass from that predicted by the mean relation, an excess of nearby galaxies from massive or less massive halos and/or substructure (as detected by the Dressler-Shectman \citep{DreShe88} test) were found relative to the population without these outliers.

The filamentary structures and mass planes, and the mass fraction in
them, provide an additional characterization of individual cluster environments.
The WCS mock observations along the 96 lines of sight of each cluster can now be compared to $|\cos\theta|$, where $\theta$ is the angle between
the line of sight and the normal to these
planes.  In addition to the normal to four of the
planes mentioned above (connected filament mass, filamentary mass, halo mass, 
and galaxy richness above $0.2L_*$), 
we also consider a fifth preferred direction, the angle to the nearest filament, and in this case use $|\sin \theta_{\rm fil}|$ of this angle (i.e. $|\cos \theta|$ of the associated normal to the nearest filament). 

The rough expectation is that a cluster's measured mass along the sheet with
the most filamentary mass or total halo mass
($|\cos\theta| \sim 0$) will be larger than that along the plane's normal vector.
This correlation is not expected to be perfect, 
as there is often a filament close to the normal, and the fraction and distribution of mass in the plane can vary.  In particular, 
the planes aren't necessarily completely filled, and some directions through this plane might not intersect large amounts of mass (i.e. there might be a lack of isotropy in the plane as mentioned earlier).
For planes which are not isotropically filled, 
one might thus expect a triangular distribution of
mass prediction (on the x-axis) vs. $|\cos \theta|$ (on the y-axis):
with low mass values for all $|\cos \theta|$, and high mass values for small
$|\cos\theta|$ (along the plane).  In addition, planes were defined only 
within 10 Mpc/h of the cluster, or less, and interlopers are expected to have an effect
sometimes at 10 times or more of that
distance.  These factors suggest that the alignment of an observational
direction with a sheet may not
be highly noticeable in observations, even if most of the local (filamentary and/or halo) mass lies within this sheet.  

\begin{figure}
\begin{center}
\resizebox{3.5in}{!}{\includegraphics{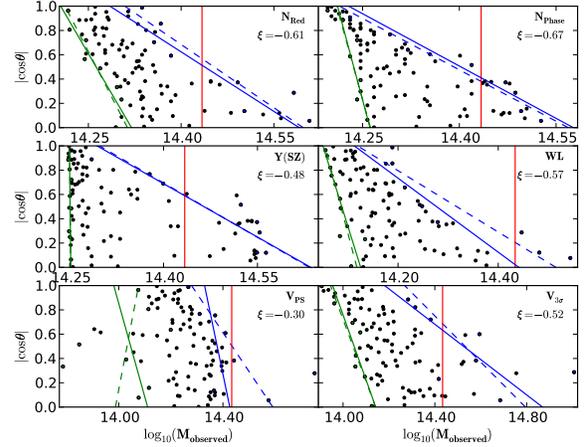}}
\end{center}
\caption{Example of mass scatter correlations:  each point is a mass measurement
for the same cluster along
one of $\sim$ 96 lines of sight, having angle $\theta$ with
the normal of the connected filament plane.   
The vertical line gives the true mass.  
The mass measurements are based upon (left to right, top to bottom):
red galaxy richness, richness based on phase space, Compton decrement, weak lensing, 
phase based velocity dispersions and 3-$\sigma$ clipping velocity dispersions.
The mass axes for each measurement vary to cover the range of masses 
found for that technique, note the scales differ.
Envelopes are fit to truncated sets of these points,
both using a chi-squared fitting (dashed line) and a shortest perpendicular
distance to the envelope (solid line), as described in the text.  Where the two
severely disagree (e.g. lower left hand box), one or both fits are bad.
The correlation coefficients between $|\cos \theta|$ and $\log_{10} M$/h
are shown at upper right, they are the smallest in absolute
value of of those for all or truncated points., or the truncated points.
The cluster 
has mass $2.7 \times 10^{14} M_\odot/h$, 9 filaments, and 
about 84\% of its connected filament mass in the connected filament plane.
}
\label{fig:onecluscorr}
\end{figure}

Even with these contraindications, for many clusters
we found a strong correlation for many mass measures with
the angle between the line of sight and the locally 
defined planes.
These strong correlations are seen not only for both
measures of richness, which in principle are closely localized to the cluster, 
but also for weak lensing, and to a lesser extent, SZ.  Correlations are both less frequent and
less strong for velocity dispersions.
We show an example of one cluster's mass scatter for the six observables
in Fig.~\ref{fig:onecluscorr}.  The measured mass is calculated using scaling from the mean mass-observable relation  for clusters in the simulation with $M \geq 10^{14} h^{-1} M_\odot$, and its value is shown versus $|\cos \theta|$, where
$\theta$ is the angle between the observational direction and the connected filament plane's normal.
The six panels show two richnesses, SZ, weak lensing, and two velocity dispersions.
This $2.7 \times 10^{14} h^{-1} M_\odot$ cluster, with 9 filaments, exhibits strong correlations for all six measurements.  It has 84\% of its
connected filament mass and 72\% of its halo mass in the connected filament plane.
%G.clusnormslist(30, 0, 1)

Given the noisiness of the data, we are mostly interested in general qualitative trends
for the full set of 226 cluster nodes.
We estimate correlations for each cluster in
two ways.  One is to use the correlation coefficient for
$(\log M,|\cos \theta|)$, 
or the truncated set of points by the procedure
described below, if that gives a lower absolute value (i.e. weaker value) for the correlation coefficient.  These are shown for our example in
Fig.~\ref{fig:onecluscorr} above.   
By eye, a correlation of $< -0.25$ appears to be a strong correlation, 
between $-0.25$ and $0.25$  is often (not always) extremely noisy, 
and a correlation $>0.25$ indicates an (unexpectedly) positive 
correlation.  We use this division hereon.  
A positive correlation is unexpected as this means that
measured cluster mass increases as the line of sight intersects less of
the preferred plane. 

\begin{figure}
\begin{center}
\resizebox{3.5in}{!}{\includegraphics{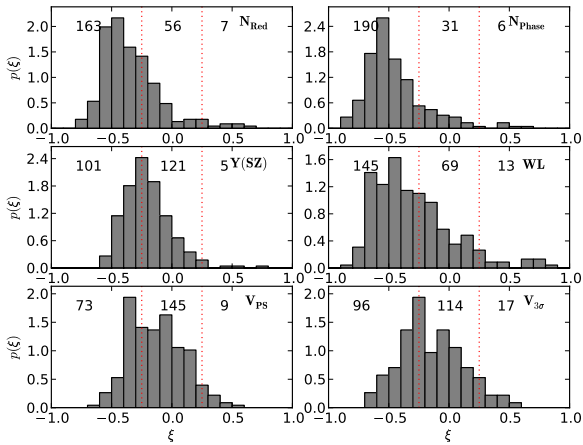}}
\end{center}
\caption{Distributions of correlations between measured mass and $|\cos \theta|$ for the 226 cluster nodes, for six observables.  Here $\theta$ is the
angle between the line of sight and the normal to the connected filament plane.  The correlation for each cluster is taken to be the one which is minimum in absolute value for all points or the truncated (as described in the text) set of points. 
The mass measurement methods are as in Fig.~\ref{fig:onecluscorr}, i.e., left, right, top to bottom are red galaxy richness, phase space richness, 
Compton decrement, weak lensing, velocity dispersion using spatial information and
3$\sigma$ clipping velocity dispersion.  Also printed for each method are
(left) the number of clusters with correlation $<-0.25$, (middle) 
the number of clusters where the correlation's absolute value is less than 0.25 (and
thus possibly noise), and (right) the number of clusters where the correlation is $>0.25$, 
i.e. both positive and large, indicating a higher mass estimate as the line of sight
becomes more perpendicular to the maximal plane.  The dashed vertical lines
separate these three regions.} 
\label{fig:obscorr}
\end{figure}

The distributions of these correlation coefficients, for the respective
mass measurements in Fig.~\ref{fig:onecluscorr} and the connected filament
plane, are shown in Fig.~\ref{fig:obscorr} for all 226 cluster nodes.
Also printed are the number of clusters with
strong (negative), noisy and positive correlations for each measurement.
The results are similar for all 5 choices of plane within the considerable noise.\footnote{
Relative to the connected filament plane shown, all other
planes have more strong negative correlations for the two richness based
masses; for planes besides the plane perpendicular to then nearest filament (which is lower), 
there are more strong correlations for weak lensing and Compton decrement, and similar numbers for velocity dispersions.  The plane perpendicular to the
nearest filament has fewer negative correlations for weak lensing and Compton decrement and many fewer for velocity dispersions. 
For all 6 mass measurements,
the median correlation for the plane perpendicular to the nearest filament is
weaker (i.e. more positive) than for the other four planes, by more than the scatter
between the median correlations for other four.
%??
%**runplotcov**
}
The fraction of clusters having strong negative or positive correlations, split according to type of mass measurement, is 
shown in table \ref{tab:corrlns}, with ranges shown for the 5 choices of plane.
The composite correlation of
$\log M_{\rm true}-\log M_{\rm pred}$ for all the clusters with $|\cos \theta|$
followed similar trends, with a strongest correlation coefficient for both richnesses, then weak lensing;
velocity dispersions and SZ are all similarly low.
%**cthetamass** in plots.sm, allslopes.
The amount of correlation between line of sight and normal to various planes
is correlated to some extent with the fraction of mass or richness 
in these planes, as might be expected.%**mpplots
There is also a correlation
between the strength of correlations of $(\log M,|\cos \theta|)$ and the alignments
between planes for each cluster (not surprisingly, this depends upon
the pair of planes being considered and the plane used to defined $\theta$).
%**dotscorrs**,**corrsourcewin**
%**plotcovs**
Considering multiwavelength measurements together for each cluster, 
$\sim$40-50\% of the clusters have a strong negative correlation (i.e. the expected
sign) for at least 3 observables.%*nneg 1** %run negslopes() first
%** plotcovs **
\begin{table*}
\begin{center}
\begin{tabular}{ccccccc}
property & red richness& phase rich & SZ & weak lensing & phase v & 3$\sigma$ v\\
 \hline
corrln $<$ -0.25   & 70-80\% & 85-90\%& 35-50\%& 55-75\% & 20-40\% & 25-40\% \\
\hline
corrln $<$ -0.25 or & 75-85\% & 85-95\%& 40-55\%& 65-75\% & 25-40\% & 30-40\% \\
neg rh slope & & & & & &\\
\hline
inertia corrln $<$-0.25  & 80-85\% & 75-80\%& 40-50\%& $>$90\% &35-45\% & 55\%  \\ 
\hline
corrln $>0.25$
&$\leq$3\%& $\leq$3\%& $\leq$2\% & 1-6\% & 1-5\%&2-7\%\\ \hline
corrln $>0.25$  or\\
large rh pos slope
&$1-4$\%& $2-5$\%& $\leq3$\% & 4-10\% & 4-9\%&2-7\%\\
\hline
ill defined slopes& 5\% & $\sim 0$&$\sim 0$ & $\sim 0$& 45-50\%&50-60\% \\
\end{tabular}
\end{center}
\caption{Cluster fractions with strongly negative (expected, rounded to the nearest 5\%) and positive (unexpected, not rounded)
correlation coefficients between $|\cos \theta|$ and measured mass, by observable,
for filament planes, 
and effect of also considering the inverse slope of
the right hand (``rh'') envelopes of the $(\log M,|\cos \theta|)$ relation, when either strongly negative (expected) or positive (unexpected).
and for directions associated with cluster inertia tensor.
The range of values encompass those for planes defined with
connected filament mass, all filamentary mass within 10 Mpc/h sphere, all halo mass within 10 Mpc/h sphere, galaxy richness, and the plane whose normal
is perpendicular to nearest filament to line of sight.
Also
shown for planes are fractions of clusters with badly defined envelopes (``ill-defined slopes''--suggesting no 
correlation).  The range for strongly negative correlations with two directions of
the inertia tensor of the cluster itself (long axis of cluster, using $\sin \theta$, or direction of eigenvector
with largest eigenvalue) is also shown, see below in text. 
}
%plotallcuts and staring at chicorr :(, fracpos_slope_corr.dat fix corr and large slope.
% get new chicorr, then plotall cuts gives some numbers, rest fracpos...etc from
%posslope, recheck corrcounts etc. **plotallcuts** 
\label{tab:corrlns}
\end{table*}
%  **runplotcov, medvals.ps has plot**

For the planes,
the correlation coefficient sometimes is low, even with a visible
trend of measured mass versus $|\cos\theta|$.  One apparent cause is
the expected triangular envelope for the points described above.
To identify this pattern, we considered slopes of approximate envelopes
of the distributions, shown in Fig.~\ref{fig:onecluscorr}.
Points are binned in 8 approximately equally filled\footnote{As mentioned earlier, lines of sight where a more massive cluster is present within $r_{180b}$ are discarded.} $|\cos \theta|$ bins, in each bin $\leq 2$
points are discarded at large or small $\log M$ if separated from their nearest neighbor by more than 6 times the median separation in mass in that bin (or the minimum separation if the median is zero).  This threw out
many of the notable outliers.  It also sometimes threw out other points, in
a binning dependent way, but the number these points is small and not
a concern as we are interested in average overall properties.
Points within $3 \sigma$ of the median  $\log M$  are
then kept within each $|\cos\theta|$ bin.  Straight line envelopes
were then fit to both ends of each bin,
either by minimizing perpendicular distance to the envelope or
minimizing the chi-squared (note of $\log M(|\cos \theta|)$).
Envelopes for both methods are shown 
in Fig.~\ref{fig:onecluscorr}, the cases shown where they strongly disagree 
correspond to one or both envelopes having bad fits.
From hereon we restrict to envelopes
based upon minimizing the perpendicular distance to
the envelope. The resulting right and left hand inverse slopes are correlated with the correlation coefficents of ($\log M,|\cos \theta|$).   We explored adding clusters
to the negative
(or positive) slope sample which have inverse slopes less than (or more than) 
the mean value of inverse slope for our correlation coefficient cutoff $\pm 0.25$;
the small effect can be seen in table \ref{tab:corrlns}.  (Sometimes the mean value
had the wrong sign, e.g. for velocity dispersions for some choice of plane, 
which have large scatter, in this case the cutoff was set to zero.)
We strove to be conservative in claiming a correlation, so that our estimates for
the strength of these planar orientational effects tend to be lower bounds.
%**plotallcutplot**

Most of the time the envelopes found by our algorithm are reasonable to the
eye, but sometimes they fail catastrophically, and were caught by the
goodness of fit estimator.  The catastrophic failures seem to occur when
no correlation is apparent between $|\cos \theta|$ and measured mass,
as does an envelope close to vertical (inverse slope close to zero).
The goodness of fits are the worst for the velocity dispersions, which have
close to half of the clusters not allowing good fits for either the left or right envelopes, 
even when the goodness of fit passes threshold, the envelopes are often close to vertical: i.e.
the minimum or maximum velocity dispersion mass is 
similar either perpendicular to the maximum plane or looking through it.
%**plotallcutplot**

For the unexpected positive correlations, 
a positive inverse left envelope slope can be understood by
looking down a filament near the perpendicular to the plane (small angle, large mass) 
and then catching a ``gap'' in the plane (large angle, small mass).
It is more difficult to understand $>0.25$ correlations or
large positive right envelope inverse slopes (i.e.  the largest measured mass
closer to the perpendicular to the dominant plane).  These do not dominate
but are not uncommon:  for any choice of plane, 
$\sim$10-20\% of %posslopevals 0, 2
the clusters have at least one observable with
strongly positive inverse right hand slope or correlation (almost
half of these are due to velocity dispersions).   This dropped to
$<5$\%  
(down to 1\% using the plane perpendicular to the nearest filament or richness) when requiring clusters to
have at least 3 observables with either right hand positive slope or correlation (most often weak lensing and both dispersion measurements).  
%**posslope**, **posslopeshort**, 
%**posslopevals**(top row) for 1 or 3 fractions
%**dotscorrs* **slopeccoeff**

Restricting to correlations, which are a cleaner and more conservative
measurement,  there are 45 clusters with a positive correlation
for at least one measurement 
(usually velocity dispersions).  These clusters differ from the full sample in
having, twice as often as the latter, high fractions of
perpendicular mass to the connected filament plane and/or some pair of planes
misaligned by 60 degrees or a recent merger (as defined earlier).   They also slightly more often have
another massive cluster within 10 Mpc/h, low mass or richness fraction in some
plane, or are more flat (as measured by its inertia tensor, smallest axis/middle axis $<0.6$).  %**weirdsource, 
Fewer than a quarter of the clusters with a positive
correlation for at least one measurement don't have one of these factors present, 
and
some of these are close to our cutoffs, e.g. have more mass within 6 Mpc/h (rather than 3 Mpc/h) to the perpendicular to
the connected filament plane than most clusters, or planes mismatching by almost
60 degrees.  
%corrsourcelist.  84, 120 big filament,
% 104 big line of things within 1/10 of mass, large mass not in conn plane 
%85 had recent major merger
%**corrsourcewin ??check 60 84 85 86 104 120 134 177
%weirdsource** in plots.sm
The ``unexplained'' strongly positive correlations occur for weak lensing and velocity dispersion mass
measurements.\footnote{
The fewest cases of strongly positive inverse slope or correlation
occur for
the plane defined using the perpendicular to the nearest filament to line of sight, 
suggesting that filaments close to the line of sight might be the cause of
positive correlations, but again positive correlations did not always occur for
these configurations.
However, the plane perpendicular to the nearest filament also gives the fewest (except for richness)
strongly negative correlations, i.e. its correlations are weaker in general.}
%**corrsourcewin**
Given the complexity of the cosmic web, and the small region we
use to characterize the cluster's environment, it is to be expected that
our simple cartoon description will not always correlate precisely
with observables.

Similar correlations can be calculated using 
two axes defined from the inertia tensor for the cluster
itself: the ``long'' axis of the cluster, and 
the normal to the direction of the eigenvector for the largest eigenvalue of its inertia tensor (pointing orthogonal to the longest 
and middle axes of the cluster).  As mentioned earlier, 
these directions are correlated with
the planes, with the  ``long'' cluster axis tending to lie within
them and the latter direction tending to align with the plane normals.
Compared to the five planes above,
the median correlation with mass scatter is stronger
for red galaxy richness, weak lensing and the two velocity dispersions, is similar for
SZ, and brackets that for phase richness (the long axis of the cluster always
has the stronger correlation of the two).   The strength of effect for the ``long'' axis of the cluster is likely due to not only a {\it plane} being compared to
the line of sight, but a specific
high density axis within that plane; almost 90\% of the clusters have a strong
negative correlation for at least 3 of the 6 observables.  The fractions of
strong negative correlations for these two directions determined by
the cluster inertia tensor is also shown in table \ref{tab:corrlns}.

In summary, 
the mass scatter for richness, Compton decrement, weak lensing and velocity
dispersion measures is often correlated with the angle to these planes (most for richnesses, and least for velocity dispersions). 
The correlations aren't perfect
and can sometimes be weak, or even of the opposite sign than expected.
In the latter case it is often also true that 
the different dominant planes (mass,  connected or all filamentary halos, and richness) are not well aligned, or that
a large filament extends perpendicular to the connected filament plane.
Besides being correlated with each other, the planes and the mass
scatters are also correlated with axes of the cluster's inertia tensor.

\section{Conclusion}
After implementing a filament finder on an N-body simulation, we studied the
resulting filamentary environment for the 226 nodes which are also
clusters ($M\geq 10^{14} M_\odot/h$).
Filaments tend to lie in sheets, presumably those from which they condensed, providing a highly
anisotropic environment for the cluster at their center.  Within a 10 Mpc/h sphere, we identified
sheets of width 3 Mpc/h, centered on each cluster, which maximize either total mass, 
connected filament mass, all filamentary mass, 
or galaxy richness.  The all filament and halo mass planes are most often closely
aligned, while the connected filament plane tends to be within the pair of least aligned
planes for the majority of clusters.
The direction of the filamentary and mass planes persist slightly as
the 10 Mpc/h spheres are extended to 20 Mpc/h.
We measured the correlation of mass
measurement scatters with the direction of observation relative to these
planes for mock observations of richness, Compton decrement, weak lensing and
velocity dispersions, via correlation coefficients and fits to the envelopes of
the measurements.  Often there is a strong correlation between measured
mass and direction to the local plane, in spite of
the relatively small region (10 Mpc/h radius) used to define the plane (again,
this correlation might be overestimated for Compton decrement and weak lensing,
which both can have strong scatter from distances larger than our box size).
Strong correlations are least likely
for velocity dispersions, and fitting envelopes to their distribution of
$|\cos\theta|$ versus $\log M$ tend to fail badly.
This is perhaps not surprising because our finder doesn't include
dynamical information.   Alignments of observational direction with two of the
axes of the inertia tensor of the cluster also results in strong correlations with
measured mass scatter.

How these planes and correlations with
scatter extends to higher redshift depends upon
how the finder extends to higher redshift.  This is a subtle
question as the finder of \citet{Zha09} has a built-in scale: 
a cutoff for minimum halo mass.
A full analysis of appropriate generalizations is beyond the scope of
this paper; two natural possibilities, however, are to leave the
minimum mass alone, or to choose a minimum mass so that the ratio of the 
number of halos 
to the number of clusters (107 at $z=0.5$, 25 at $z=1.0$) remains the same,
which gives a minimum mass
of $8.2 \times 10^{10} h^{-1} M_\odot$ for $z=0.5$ and $3.0\times 10^{11} h^{-1} M_\odot$ for $z=1.0$.\footnote{We thank M. White for this suggestion.} 
Choosing the latter case (and luminosity cut at $0.2L^*$), most of the
trends persist to these higher redshifts, although the total number of filaments
in the box decreases.  For $z=0.5$ and $z=1.0$, the planar mass fractions around clusters
are close to unchanged.  
For all 3 redshifts, there is a slight drop in richness fraction in the richness plane as redshift increases, and 
the halo mass fraction in planes around random points appears to grow, so that by $z=1.0$ it is comparable to that around the 25 clusters in the box at
$z=1.0$.%
%**threezplanes
For correlations of plane directions with cluster observations, the statistics are very noisy for $z=1.0$, but
for $z=0.5$, the fractions of clusters with strong (expected) negative
correlations of angle with plane and mass scatter\footnote{The model for color assignments
is valid only for $z=0.1$, so we did not
consider red galaxy richness at other $z$.}, as in table \ref{tab:corrlns}, tend to either remain the same in range or
slightly increase (velocity dispersions do decrease in one case), the number of
clusters with at least three negative correlations are close to unchanged for three
planes,
dropping for the richness and nearest filament planes,
% **nneg** 1 0.6760
and positive
correlation fractions are about the same except for (an increase for) velocity dispersions.
Large plane misalignments are less common, but clusters with
misaligned planes still are more likely to have smaller mass fractions in the 
plane or more perpendicular mass than the full sample.   

It would be interesting to determine whether this generalization to
higher redshift is appropriate and then to understand the results
in terms of the evolution of the filamentary neighborhood of the clusters and
the clusters within them.

The correlations between mass scatter and angle of observation with the planes
(and inertia tensor of the cluster) rely upon three dimensional information
available to us as simulators.  It would be very interesting to find a way
to make this source of mass bias more evident to observers, perhaps by using
a filament finder based upon galaxies directly (amongst those mentioned earlier), and
seeing how well they trace these planes, or by combining multiwavelength measurements.
In depth studies underway of cluster environments such as
\citet{Lub09} would be excellent datasets to apply and refine such methods.

We thank Y. Birnboim, O. Hahn, S. Ho, G. Jungman, D. Keres, L. Lubin, S. Nagel, S. Shandarin, A. Szalay, D. Zaritsky and
especially M. White for helpful discussions, and we thank M. George, L. Lubin and E. Rozo for helpful
comments on the draft.  JDC thanks R. Sheth for
an introduction to research on filaments, and
the Aspen Center for Physics for
providing the place and opportunity for us to meet and discuss.
YN thanks the Santa Fe Cosmology School
for the opportunity to present this work, and we both thank the participants there for many
useful comments and questions, and S. Habib and K. Heitmann for support in order to be
able to attend.
Martin White's simulations, used in this paper,  
were performed at the National Energy
Research Scientific Computing Center and the Laboratory Research Computing
project at Lawrence Berkeley National Laboratory. 


\begin{thebibliography}{99.}

\bibitem[{{Abell}(1958)}]{Abe58}
Abell G.O., 1958, ApJS, 3, 211

\bibitem[{{Altay, Colberg \& Croft}(2006)}]{AltColCro06}
 Altay, G., Colberg, J.~M., 
\& Croft, R.~A.~C.\ 
%{\sl The influence of large-scale structures on halo shapes and alignments }, 
2006, MNRAS, 370, 1422 

\bibitem[{{Aragon-Calvo et al.}(2007a)}]{Ara07a} 
Arag{\'o}n-Calvo, M.~A., Jones, B.~J.~T., van de Weygaert, R., 
\& van der Hulst, J.~M., 
% {\sl The multiscale morphology filter: identifying and extracting spatial patterns in the galaxy distribution }, 
2007a, A \& A, 474, 315 

\bibitem[{{Arag{\'o}n-Calvo et al.}(2007b)}]{Ara07b}  
Arag{\'o}n-Calvo, M.~A., van de Weygaert, R., Jones, B.~J.~T., 
\& van der Hulst, J.~M., 
%{\sl Spin Alignment of Dark Matter Halos in Filaments and Walls }, 
2007b, ApJL, 655, L5 

\bibitem[{{Arag{\'o}n-Calvo, Shandarin \& Szalay}(2010)}]{AraShaSza10}  
Arag{\'o}n-Calvo, M.A., Shandarin, S.F., Szalay, A., 2010, arXiv:1006.4178

\bibitem[{{Arag{\'o}n-Calvo, van de Weygaert \& Jones}(2010)}]{AravanJon10}
Arag{\'o}n-Calvo, M.A., van de Weygaert, R., Jones, B.J.T., 2010, arxiv:1007.0742
%Multiscale Phenomenology of the Cosmic Web

\bibitem[{{Bailin \& Steinmetz}(2005)}]{BaiSte05}
 Bailin, J., \& Steinmetz, M., 
%{\sl Internal and External Alignment of the Shapes and Angular Momenta of {$\Lambda$}CDM Halos }, 
2005, ApJ, 627, 647 

\bibitem[{{Barrow, Bhavsar \& Sonoda}(1985)}]{BarBhaSon85} 
Barrow, J.~D., Bhavsar, S.~P., 
\& Sonoda, D.~H., 
%{\sl Minimal spanning trees, filaments and galaxy clustering }, 
1985, MNRAS, 216, 17 

\bibitem[{{Basilakos et al}(2006)}]{Bas06}
Basilakos, S.; Plionis, M.; Yepes, G.; Gottlöber, S.; Turchaninov, V., 2006, MNRAS, 365, 539 %{\sl "The shape-alignment relation in $\Lambda$cold dark matter cosmic structures"} --superclusters

\bibitem[{{Becker \& Kravtsov}(2010)}]{BecKra10}
Becker, M.R., Kravtsov, A.V., 2010, arXiv:1011.1681

\bibitem[{{ Betancort-Rijo \&  Trujillo}(2009)}]{BetTru09} 
Betancort-Rijo, J.E., Trujillo, I., 
%{\sl An analytical framework to describe the orientation of dark matter halos and galaxies within the large-scale structure}, 
2009, arXiv:0912.1051

\bibitem[{{Bharadway et al.}(2000)}]{Bha00}
 Bharadway, S., Sahni, V., Sathyaprakash, B.S., Shadarin, S.F., 
%{\sl Evidence for Filamentarity in the Las Campanas Redshift Survey}, 
2000, ApJ 528, 21

\bibitem[{{Biviano et al.}(2006)}]{Biv06} 
Biviano, A., Murante, G., Borgani, S., Diaferio, A., Dolag, K., Girardi, M., 
%{\sl On the efficiency and reliability of cluster mass estimates based on member galaxies}, 
2006, A\& A, 456, 23

\bibitem[{{Bond, Strauss \& Cen}(2009)}]{BonStrCen09} 
Bond, N., Strauss, M., Cen, R., 
%{\sl Crawling the Cosmic Network: Exploring the Morphology of Structure in the Galaxy Distribution}, 
2009, arxiv:0903.3601

\bibitem[{{Bond, Strauss \& Cen}(2010)}]{BonStrCen10} 
 Bond, N.A., Strauss, M.A., Cen, R., 
%{\sl Crawling the Cosmic Network: Identifying and Quantifying Filamentary Structure}, 
 2010, arXiv:1003.3237

\bibitem[{{Bond, Kofman \& Pogosyan}(1996)}]{BonKofPog96} 
Bond J. R., Kofman L., Pogosyan D.,  
%{\sl How filaments of galaxies are woven into the cosmic web}, 
1996, Nature, 380, 603

\bibitem[{{Cen}(1997)}]{Cen97}
Cen R., 1997, ApJ, 485, 39

\bibitem[{{Chambers, Melott \& Miller}(2000)}]{ChaMelMil00} 
Chambers, S.W., Melott, A.L., Miller, C.J., 
%{\sl Einstein Cluster Alignments Revisited}, 
2000, ApJ, 544, 104

\bibitem[{{Choi et al.}(2010)}]{Cho10}
Choi, E., Bond, N.A., Strauss, M.A., Coil, A.L., Davis, M., Willmer, C.N.A., 2010, MNRAS, 
406, 320

\bibitem[{{Colberg et al.}(1999)}]{Col99} 
Colberg, J., White, S.D.M., Jenkins, A., Pearce, F.R., 
%{\sl Linking Cluster formation to large scale structure}, 
1999, MNRAS, 308, 593 

\bibitem[{{Colberg, Krughoff \& Connolly}(2005)}]{ColKruCon05}
Colberg, J., Krughoff, K.S., Connolly, A.J., 2005, MNRAS, 359, 272

\bibitem[{{Colberg}(2007)}]{Col07} 
Colberg, J.~M., 
%{\sl Quantifying cosmic superstructures }, 
2007, 
MNRAS, 375, 337 

\bibitem[{{Colombi, Pogosyan \& Souradeep}(2000)}]{ColPogSou00} 
Colombi, S., Pogosyan, D., Souradeep, T., 
%{\sl Tree Structure of a Percolating Universe}, 
2000, PRL 85, 5515

\bibitem[{{Conroy, Gunn \& White}(2009)}]{SP1}
Conroy C., Gunn J.E., White M., 2009, ApJ, 699, 486
\bibitem[{{Conroy, White, \& Gunn}(2010)}]{SP2}
Conroy C., White M., Gunn J.E., 2010, ApJ, 708, 58
\bibitem[{{Conroy, \& Gunn}(2010)}]{SP3}
Conroy C., Gunn J.E., 2010, ApJ, 712, 833

\bibitem[{{Costa-Duarte, Sodre, \& Durret}(2010)}]{CosSodDur10}
%{\sl Morphological Properties of Superclusters of Galaxies}, 
Costa-Duarte, M.V., Sodre Jr., L., Durret, F., 2010, 
arXiv:1010.0981 

\bibitem[{{Dalton et al}(1992)}]{Dal92}
Dalton, G.B., Efstathiou, G., Maddox, S.J., \& Sutherland, W.J., 1992,
ApJL 390, L1

\bibitem[{{Davis et al.}(1985)}]{DEFW}
Davis M., Efstathiou G., Frenk C.S., White S.D.M., 1985, ApJ, 292, 371

\bibitem[{{Dekel et al.}(2009)}]{Dek09}
Dekel, A., et al., 
2009, Nature, 457, 451 

\bibitem[{{den Hartog \& Katgert}(1996)}]{denHartog}
den Hartog, R., Katgert, P., 1996, MNRAS, 279, 349

\bibitem[{{Diaferio \& Geller}(1997)}]{DiaGel97}
Diaferio, A., Geller, M.J., 1997, ApJ, 481, 633

\bibitem[{{ Diemand, Kuhlen \& Madau}(2006)}]{DieKuhMad06}
 Diemand J., Kuhlen M., Madau P., 
%{\sl Early supersymmetric cold dark matter substructure}, 
2006, ApJ, 649, 1

\bibitem[{{Dolag et al.}(2006)}]{Dol06} 
Dolag, K., Meneghetti, M., Moscardini, L., Rasia, E., 
\& Bonaldi, A., 
%{\sl Simulating the physical properties of dark matter and gas inside the cosmic web }, 
2006, MNRAS, 370, 656 

\bibitem[{{Dressler \& Shectman}(1988)}]{DreShe88}
Dressler A., Shectman S.A., 1988, AJ, 95, 985

\bibitem[{{Einasto et al.}(1984)}]{Ein84}
Einasto, J., Klypin, A.A., Saar, E., Shandarin, S.F., 1984, MNRAS, 206, 529

\bibitem[{{Faltenbacher et al.}(2002)}]{Fal02} 
Faltenbacher, A., Gottl{\"o}ber, S., Kerscher, M., 
Muller, V., 
%{\sl Correlations in the orientations of galaxy clusters }, 
2002, A \& A, 395, 1 

\bibitem[{{Faltenbacher et al.}(2005)}]{Fal05} 
Faltenbacher, A., Allgood, B., Gottl{\"o}ber, S., Yepes, G., 
\& Hoffman, Y., %{\sl Imprints of mass accretion on properties of galaxy clusters }, 
 2005, MNRAS, 362, 1099 

\bibitem[{{Faltenbacher et al.}(2007)}]{Fal07} 
Faltenbacher, A., Li, C., Mao, S., van den Bosch, F.C., Yang, X., Jing, Y.P., Pasquali, A., Mo, H.J., 
%{\sl Three Different Types of Galaxy Alignment within Dark Matter Halos}, 
2007, ApJL, 662, 71

\bibitem[{{Feix et al.}(2008)}]{Fei08}
Feix, M., Xu, D., Shan, H., Famaey, B., Limousin, M., Zhao, H., 
\& Taylor, A., 
% {\sl Is Gravitational Lensing by Intercluster Filaments Always Negligible? }, 
2008, ApJ, 682, 711 


\bibitem[{{Forero-Romero et al.}(2009)}]{For09} 
Forero-Romero J.E., Hoffman Y., Gottlober S., Klypin, A., 
Yepes G., 
% {\sl A dynamical classification of the cosmic web }, 
 2009, MNRAS, 396, 1815

\bibitem[{{Gal et al.}(2008)}]{Gal08} 
Gal, R.R., Lemaux, B.C., Lubin, L.M., Kocevski, D., Squires, G.K., 
%{\sl The Complex Structure of the Cl 1604 Supercluster at z $\sim$ 0.9}, 
2008, ApJ, 684, 933


\bibitem[{{Gay et al.}(2009)}]{Gay09}
Gay, C., Pichon, C., Le Borgne, D., Teyssier, R., Sousbie, T., 
\& Devriendt, J., 
% {\sl On the filamentary environment of galaxies }, 
2009, arXiv:0910.1728  %(colors)

\bibitem[{{Genovese et al.}(2010)}]{Gen10}
Genovese, C.R., Perone-Pacifico, M., Verdinelli, I., Wasserman, L., 
%{\sl Nonparametric Filament Estimation}, 
2010, arXiv:1003.5536

\bibitem[{{Gonzalez \& Padilla}(2009)}]{GonPad09} 
Gonzalez, R.E., Padilla, N., 
%{\sl Automated detection of filaments in the large scale structure of the universe}, 
 2009, arxiv:0912.0006, MNRAS to appear

\bibitem[{{Hahn et al.}(2007a)}]{Hah07a} 
Hahn, O., Porciani, C., Carollo, C. M., Dekel, A., 
%{\sl Properties of dark matter haloes in clusters, filaments, sheets and voids}, 
2007a, MNRAS, 375, 489

\bibitem[{{Hahn et al.}(2007b)}]{Hah07b} 
Hahn, O., Carollo, C. M., Poricani, C., Dekel, A., 
%{\sl The Evolution of Dark Matter Halo Properties in Clusters, Filaments, Sheets and Voids}, 
2007b, MNRAS, 381, 4


\bibitem[{{Hahn, Teyssier \& Carollo}(2010)}]{HahTeyCar10} 
Hahn, O., Teyssier, R., Carollo, C.M., 
%{\sl The large-scale orientation of disc galaxies}, 
2010, arXiv:1002.1964, MNRAS to appear

\bibitem[{{Hallman et al}(2007)}]{Hal07}
Hallman, E.J., O'Shea, B.W., Burns, J.O., Norman, M.L., Harkness, R., Wagner, 
R., 2007, ApJ 671, 27

\bibitem[{{Hoekstra}(2001)}]{Hoe01}
Hoekstra H., 2001, A\&A, 370, 743

\bibitem[{{Holder, McCarthy \& Babul}(2007)}]{HolMcCBab07}
Holder, G.P., McCarthy, I.G., Babul, A., 2007, MNRAS 382, 1697

\bibitem[{{Hopkins, Bahcall \& Bode}(2004)}]{HopBahBod04}
Hopkins, P.F., Bahcall, N., Bode, N., 2004, ApJ 618, 1
%Cluster Alignments and Ellipticities in LCDM Cosmology 

\bibitem[{{Jing \& Suto}(2002)}]{JinSut02}
Jing, Y.P., Suto, Y., 2002, ApJ, 574, 538
%cluster ellipticity

\bibitem[{{Jones, van de Weygaert \& Arag{\'o}n-Calvo}(2010)}]{JonvanAra10} 
Jones, B.J.T., van de Weygaert, R., Arag{\'o}n-Calvo, M.A., 
%{\sl Fossil evidence for spin alignment of SDSS galaxies in filaments}, 
arXiv:1001.4479

\bibitem[{{Kartaltepe et al.}(2008)}]{Kar08} 
Kartaltepe, J.~S., Ebeling, H., Ma, C.~J., 
\& Donovan, D., 
%{\sl Probing the large-scale structure around the most distant galaxy clusters from the massive cluster survey }, 
2008, MNRAS, 389, 1240 

\bibitem[{{Kasun \& Evrard}(2005)}]{KasEvr05}
Kasun S.F., Evrard A.E., 2005, ApJ, 629, 781

\bibitem[{{Keres et al.}(2005)}]{Ker05} 
Kere{\v s}, D., Katz,  
Weinberg, D.~H.\, \& Dav{\'e}, R., 
%{\sl How do galaxies get their gas?}
 2005, MNRAS, 363, 2 


\bibitem[{{Keres et al.}(2009)}]{Ker09} 
Kere{\v s}, D., Katz, N., Fardal, M., Dav{\'e}, R., 
\& Weinberg, D.~H., 
%{\sl Galaxies in a simulated {$\Lambda$}CDM Universe - I. Cold mode and hot cores }, 
 2009, MNRAS, 395, 160 


\bibitem[{{Koester et al.}(2007)}]{maxBCG}
Koester B.P., et al., 2007, ApJ, 660, 221

\bibitem[{{Lee}(2004)}]{Lee04} 
Lee, J., 
%{\sl The Intrinsic Inclination of Galaxies Embedded in 
%Cosmic Sheets and Its Cosmological Implications: An Analytic Calculation }, 
2004, ApJL, 614, L1 

\bibitem[{{Lee}(2006)}]{Lee06}
Lee, J., 2006, astro-ph/0605697 %--ps description of filament collapse

\bibitem[{{Lee \& Evrard}(2007)}]{LeeEvr07} 
Lee, J., Evrard, A.E., 
%{\sl Cluster-Supercluster Alignments}, 
2007, ApJ 657, 30

\bibitem[{{Lee et al.}(2008)}]{Lee08} 
Lee, J., Springel, V., Pen, U.-L., Lemson, G., 
%{\sl Quantifying the Cosmic Web I: The large-scale halo ellipticity-ellipticity and ellipticity-direction correlations}, 
2008, MNRAS 389, 1266

\bibitem[{{Lee \& Springel}(2009)}]{LeeSpr09} 
Lee, J., Springel, V., 
%{\sl A Scaling Relation of the Evolving Tidal Fields in a LCDM Cosmology}, 
2009, arXiv:0911.4755, JCAP to appear


\bibitem[{{Lubin et al.}(2009)}]{Lub09} 
Lubin, L.M., Gal, R.R., Lemaux, B.C., Kocevski, D.D., Squires, G.K., 
%{\sl The Observations of Redshift Evolution in Large Scale Environments (ORELSE) Survey : I. The Survey Design and First Results on Cl 0023+0423 at z = 0.84 and RX J1821.6+6827 at z = 0.82}, 
2009, AJ, 137, 4867


\bibitem[{{Lumsden et al}(1992)}]{Lum92}
Lumsden, S.L., Nichol, R.C., Collins, C.A., Guzzo, L., 1992, MNRAS 258,1

\bibitem[{{Mead, King \& McCarthy}(2010)}]{MeaKinMcC10} 
Mead, J.~M.~G., King, L.~J., 
\& McCarthy, I.~G., 
%{\sl Probing the cosmic web: inter-cluster filament detection using gravitational lensing }, 
2010, MNRAS, 401, 2257

\bibitem[{{Mecke, Buchert \& Wagner}(1994)}]{MecBucWag94}
Mecke, K.~R., Buchert, T., 
\& Wagner, H., 
%{\sl Robust morphological measures for large-scale structure in the %Universe 
 1994, A \& A, 288, 697 

\bibitem[{{Meneghetti et al.}(2010)}]{Men10}
Meneghetti, M., Fedeli, C., Pace, F., Gottloeber, S., Yepes, G., 2010, arXiv:1003.4544

\bibitem[{{Metzler, White \& Loken}(2001)}]{MetWhiLok01}
Metzler C., White M., Loken C., 2001, ApJ, 547, 560

\bibitem[{{Murphy, Eke \& Frenk}(2010)}]{MurEkeFre10}
Murphy, D.N.A., Eke, V.R., Frenk, C.S., 2010, arXiv:1010:2202
%Filamentary structure in the 2dFGRS 

\bibitem[{{Novikov, Colombi \& Dore}(2006)}]{NovColDor06} 
Novikov, D., Colombi, S., 
\& Dor{\'e}, O., 
%{\sl Skeleton as a probe of the cosmic web: the two-dimensional case }, 
2006, MNRAS, 366, 1201 

\bibitem[{{Onuora \& Thomas}(2000)}]{OnuTho00} 
Onuora, L.I., Thomas, P.A., 
%{\sl The Alignment of Clusters using Large Scale Simulations}, 
2000, MNRAS, 319, 614

\bibitem[{{Pandey \& Somnath}(2006)}]{PanSom06} 
Pandey, B., Somnath, B., 
%{\sl The Luminosity, Colour and Morphology dependence of galaxy filaments in the Sloan Digital Sky Survey Data Release Four}, 
2006, MNRAS, 372, 827


\bibitem[{{Paz, Stasyszyn \& Padilla}(2008)}]{PazStaPad08} 
Paz, D.~J., Stasyszyn, F., 
\& Padilla, N.~D., 
%{\sl Angular momentum-large-scale structure alignments in {$\Lambda$}CDM models and the SDSS }, 
2008, MNRAS, 389, 1127 


\bibitem[{{Pereira, Bryan \& Gill}(2008)}]{PerBryGil08} 
Pereira, M.~J., Bryan, G.~L., 
\& Gill, S.~P.~D., 
%{\sl Radial Alignment in Simulated Clusters }, 
2008, ApJ, 672, 825 


\bibitem[{{Pimbblet}(2005a)}]{Pim05a} 
Pimbblet, K.~A., 
%{\sl A new algorithm for the detection of 
%intercluster galaxy filaments using galaxy orientation alignments }, 
2005a, MNRAS, 358, 256 

\bibitem[{{Pimbblet}(2005b)}]{Pim05b} 
Pimbblet, K.~A., 
 %{\sl Pulling Out Threads from the Cosmic Tapestry: Defining Filaments of Galaxies }, 
 2005b, Publications of the Astronomical 
Society of Australia, 22, 136 


\bibitem[{{Pimbblet, Drinkwater \& Hawkrigg}(2004)}]{PimDriHaw04} 
Pimbblet, K.~A., Drinkwater, M.~J., 
\& Hawkrigg, M.~C., 
%{\sl Intercluster filaments of galaxies programme: abundance and distribution of filaments in the 2dFGRS catalogue }, 
2004, MNRAS, 354, L61 

\bibitem[{{Poggianti et al.}(2010)}]{Pog10}
Poggianti, B.M., De Lucia, G., Varela, J., Arag{\'o}n-Salamanca, A., Finn, R., Desai, V., von der Linden, A., White, S.D.M., 2010, arxiv:1002.3465
%projected local density distribution for clusters as a function of redshift

\bibitem[{{Pogosyan et al.}(2009)}]{Pog09} 
Pogosyan, D., Pichon, C., Gay, C., Prunet, S., Cardoso, J.~F., 
Sousbie, T., 
\& Colombi, S., 
 %{\sl The local theory of the cosmic skeleton }, 
2009, MNRAS, 396, 635 


\bibitem[{{Porter \& Raychaudhury}(2005)}]{PorRay05}
 Porter, S.~C., 
\& Raychaudhury, S., 
% {\sl The Pisces-Cetus supercluster: a remarkable filament of galaxies in the 2dF Galaxy Redshift and Sloan Digital Sky surveys }, 
2005, MNRAS, 364, 1387  


\bibitem[{{de Putter \& White}(2005)}]{PutWhi05}
de Putter R., White M., 2005, New Astronomy, 10, 676

\bibitem[{{Ragone-Figueroa \& Plionis}(2007)}]{RagPli07}
Ragone-Figueroa, Cinthia; Plionis, Manolis, 2007, MNRAS, 377, 1785 
%{\sl "Environmental influences on the morphology and dynamics of group-sized haloes"}


\bibitem[{{Reblinsky \& Bartelmann}(1999)}]{RebBar99}
Reblinsky K., Bartelmann M., 1999, A\&A, 345, 1

\bibitem[{{Sahni, Sathyaprakash \& Shandarin}(1998)}]{SahSatSha98} 
Sahni, V., Sathyaprakash, B.~S., 
\& Shandarin, S.~F., 
%{\sl Shapefinders: A New Shape Diagnostic for Large-Scale Structure }, 
1998, ApJL, 495, L5 


\bibitem[{{Schaefer}(2009)}]{Sch09} 
Sch{\"a}fer, B.~M., 
%{\sl Galactic Angular Momenta and Angular Momentum Correlations in the Cosmological Large-Scale Structure }, 
2009, 
International Journal of Modern Physics D, 18, 173 

\bibitem[{{Schmazling}(1998)}]{Sch98} 
Schmalzing, J., , 1998,  Proceedings of the 12th Potsdam Cosmology Workshop (1997). Eds. V.Mueller, S.Gottloeber, J.P.Muecket, J.Wambsganss, 195ff

\bibitem[{{Schmalzing et al.}(1999)}]{Sch99} 
Schmalzing, J., Buchert, T., Melott, A.~L., Sahni, V., Sathyaprakash, 
B.~S., 
\& Shandarin, S.~F., 
%{\sl Disentangling the Cosmic Web. I. Morphology of Isodensity Contours }, 
1999, ApJ, 526, 568 

\bibitem[{{Shandarin \& Zel'dovich}(1983)}]{ShaZel83}
Shandarin, S.F., Zel'dovich, Ia. B., 1983, Comments on Astrophysics, 10, 33

\bibitem[{{Shandarin}(2004)}]{Sha04} 
Shadarin, S.F., 
%{\sl Morphological Statistics of the Cosmic Web }, 
2004, astro-ph/0405303

\bibitem[{{Shadarin, Sheth \& Sahni}(2004)}]{ShaSheSah04}
Shandarin, S.F., Sheth, J.V., Sahni, V., 2004, 
MNRAS, 353, 162
%morph of supercluster-void network in lcdm cosmology


\bibitem[{{Shandarin, Habib \& Heitmann}(2009)}]{ShaHabHei09} 
Shandarin, S.~F., Habib, S., Heitmann, K., 
%{\sl Origin of the Cosmic Network: Nature vs. Nurture}, 
2009, arXiv: 0912.4471

\bibitem[{{Shandarin}(2010)}]{Sha10}
Shandarin, S., 2010, arXiv:1011.1924

\bibitem[{{Shaw, Holder \& Bode}(2008)}]{ShaHolBod08}
Shaw, L.D., Holder, G.P., Bode, P., 2008, ApJ, 686, 206

\bibitem[{{Shen et al.}(2006)}]{She06} 
Shen, J., Abel, T., Mo, H.~J., 
\& Sheth, R.~K., 
%{\sl An Excursion Set Model of the Cosmic Web: The Abundance of Sheets, Filaments, and Halos }, 
2006, ApJ, 645, 783

\bibitem[{{Sheth et al.}(2003)}]{She03}
Sheth, J.V., Sahni, V., Shandarin, S.F. \& Sathyaprakash, B., 2003, MNRAS, 343, 22
%methods paper

\bibitem[{{De Simone, Maggiore, \& Riotto}(2010)}]{SimMagRio10}
De Simone, A., Maggiore, M., Riotto, A., 2010, arXiv:1007.1903

\bibitem[{{Skibba \& Sheth}(2009)}]{SkiShe09} 
Skibba, R. A., Sheth, R.K., %{\sl A Halo Model of Galaxy Colors and Clustering in the SDSS}, 
2009, MNRAS 392, 1080 

\bibitem[{{Song \& Lee}(2010)}]{SonLee10}
Song, H, Lee, J., 2010, arXiv:1006.4101  %ps for halo-filament mergers

\bibitem[{{Sousbie et al.}(2008a)}]{Sou08a} 
Sousbie, T., Pichon, C., Courtois, H., Colombi, S., 
\& Novikov, D., 
%{\sl The Three-dimensional Skeleton of the SDSS }, 
2008a, ApJL, 672, L1 

\bibitem[{{Sousbie et al.}(2008b)}]{Sou08b} 
Sousbie, T., Pichon, C., Colombi, S., Novikov, D., 
\& Pogosyan, D., 
%{\sl The 3D skeleton: tracing the filamentary structure of the Universe }, 
2008b, MNRAS, 383, 1655 


\bibitem[{{Sousbie, Colombi \& Pichon}(2009)}]{SouColPic09} 
Sousbie, T., Colombi, S., 
\& Pichon, C., 
%{\sl The fully connected N-dimensional skeleton: probing the evolution of the cosmic web }, 
2009, MNRAS, 393, 457 

\bibitem[{{Sousbie}(2010)}]{Sou10}
Sousbie, T., 2010, %{\sl The persistent cosmic web and its filamentary structure I: Theory and implementation }, 
 arXiv:1009.4015

\bibitem[{{Sousbie, Pichon \& Kawahara}(2010)}]{SouPicKaw10}
Sousbie, T., Pichon, C., Kawahara, H., 2010, arXiv:1009.4014
%applying to catalogues, finding cluster this way.


\bibitem[{{Stoica et al.}(2005)}]{Sto05} 
Stoica, R.~S., Mart{\'{\i}}nez, V.~J., Mateu, J., 
\& Saar, E., 
%{\sl Detection of cosmic filaments using the Candy model }, 
2005, A \& A, 434, 423 

\bibitem[{{Stoica, Martinez \& Saar}(2008)}]{StoMarSaa08} 
Stoica, R.~S., Martinez, V.~J., 
\& Saar, E., 
%{\sl A three dimensional object point process for detection of cosmic filaments }, 
 2008, arXiv:0809.4358 

\bibitem[{{Stoica, Martinez \& Saar}(2009)}]{StoMarSaa09} 
 Stoica, R.S., Martinez, V.J., Saar, E., 
% {\sl Filaments in observed and mock galaxy catalogues}, 
2009, arXiv: 0912.2021

\bibitem[{{Splinter et al.}(1997)}]{Spl97}
Splinter, R.J., Melott, A. L., Linn, A., Buck, C., Tinker, J., 1997, ApJ, 479, 632

\bibitem[{{Sunyaev \& Zel'dovich}(1972)}]{SunZel72} 
Sunyaev R. A. \& Zel'dovich, Ya. B., 1972, 
%{\sl The Observations of Relic Radiation as a Test of the Nature of X-Ray
%Radiation from the Clusters of Galaxies}, 
1972,  Comm. Astrophys, Space Phy., 
4, 173

\bibitem[{{Sunyaev \& Zel'dovich}(1980)}]{SunZel80} 
%{\sl Microwave background radiation as a probe of the contemporary structure and history of the universe}, 
Sunyaev R. A. \& Zel'dovich, Ya. B., 
1980,  ARA\&A, 18, 537

\bibitem[{{Tanaka et al.}(2009)}]{Tan09} 
Tanaka, M., Finoguenov, A., Kodama, T., Koyama, Y., Maughan, B., 
\& Nakata, F., 
% {\sl The spectroscopically confirmed huge cosmic structure at z = 0.55 }, 
2009, A \& A, 505, L9  %example again

\bibitem[{{Tormen}(1997)}]{Tor97}
Tormen G., 1997, MNRAS, 290, 411 

\bibitem[{{van Haarlem \& van de Weygaert}(1993)}]{vanvan93}
van Haarlem, M., van de Weygaert, R., 1993, ApJ, 418, 544

\bibitem[{{van Haarlem et al.}(1997)}]{vHaarlem97}
van Haarlem M.P., Frenk C.S., White S.D.M, 1997, MNRAS, 287, 817


\bibitem[{{van de Weygaert \& Bertschinger}(1996)}]{vanBer96}
van de Weygaert, R., Bertschinger, E., 1996, MNRAS, 281, 84

\bibitem[{{van de Weygaert}(2002)}]{van02}
van de Weygaert, R., 2002,  2002, Proceedings 2nd Hellenic Cos-
mology Workshop, 0, 153ff

\bibitem[{{van de Weygaert}(2006)}]{van06} 
van de Weygaert, R., 
 %{\sl Clusters and the Cosmic Web }, 
2006, 
arXiv:astro-ph/0607539 

\bibitem[{{van de Weygaert \& Schaap}(2007)}]{vanSch07} 
van de Weygaert, R., 
\& Schaap, W., 
% {\sl The Cosmic Web: Geometric Analysis }, 
2007, arXiv:0708.1441 

\bibitem[{{van de Weygaert et al.}(2010)}]{Wey10}
van de Weygaert, R., Vegter, G., Platen, E., Eldering, B., Kruithof, N., 2010, arXiv:1006.2765 %Alpha Shape Topology of the Cosmic Web

\bibitem[{{Voit}(2005)}]{Voi05}
Voit, G.M., 2005, Rev. Mod. Phys. 77, 207

\bibitem[{{Wang et al.}(2009)}]{Wan09} 
Wang, H., Mo, H.~J., Jing, Y.~P., Guo, Y., van den Bosch, F.~C., 
\& Yang, X., 
% {\sl Reconstructing the cosmic density field with the distribution of dark matter haloes }, 
2009, MNRAS, 394, 398

\bibitem[{{Wang et al}(2010)}]{Wan10}
Wang, H., Mo, H.J., Jing, Y.P., Yang, X., Wang, Y., 2010, arXiv:1007.0612

\bibitem[{{Way, Gazis \& Scargle}(2010)}]{WayGazSca10}
Way, M.J., Gazis, P.R., Scargle, J.D., 2010, arxiv:1009.0387

\bibitem[{{White et al}(1999)}]{Whi99}
White, R., et al 1999, AJ 118, 2014

\bibitem[{{White}(2001)}]{Whi01}
White M., 2001, A\&A, 367, 27

\bibitem[{{White, Hernquist \& Springel}(2002)}]{WhiHerSpr02}
White, M., Hernquist, L., Springel, V., 2002, ApJ, 579, 16

\bibitem[{{White}(2002)}]{TreePM}
White M., 2002, ApJS, 143, 241

\bibitem[{{White, Cohn \& Smit}(2010)}]{WCS}
White, M., Cohn, J.D., Smit, R., 
2010, arXiv:1005.3022, MNRAS to appear

\bibitem[{{Wojtak et al.}(2007)}]{Woj07}
Wojtak, R., Lokas, E.L., Mamon, G.A., Gottlober, S., Prada F., Moles M., 2007, 
 A\&A 466, 437

\bibitem[{{Wray et al.}(2006)}]{Wra06}
Wray, J.J., Bahcall, N.A., Bode, P., Boettiger, C., Hopkins, P.F., 2006, 
ApJ 652, 907 %The Shape, Multiplicity, and Evolution of Superclusters in LambdaCDM Cosmology

\bibitem[{{Wu, Batuski, \& Khalil}(2009)}]{Wu09}
Wu, Y., Batuski, D.J., Khalil, A., 2009, ApJ 707, 1160

\bibitem[{{Yang, Mo \& van den Bosch}(2008)}]{YanMovan08} 
Yang X., Mo H.J., van den Bosch F.C., 
2008, ApJ, 676, 248

\bibitem[{{Zel'dovich, Einasto \& Shandarin}(1982)}]{ZelEinSha82}
Zel'dovich, Ia.B., Einasto, J., Shandarin, S.F., 1982, Nature, 300, 407

\bibitem[{{Zhang et al.}(2009)}]{Zha09}
Zhang, Y., Yang, X., Faltenbacher, A., Springel, V., Lin, W., Wang, H., 2009, ApJ, 706, 747

\end{thebibliography}
\end{document}